\documentclass{basi}
\usepackage[T1]{fontenc}
\usepackage[british]{babel}
\usepackage[varg]{txfonts}
\usepackage{graphicx,url}
\begin{document}

\title[Astrophysical Coronae]{Astrophysical Coronae: Lessons from Modeling of the Intracluster Medium}

\author[P. Sharma]{Prateek Sharma$^1$\thanks{Department of Physics \& Joint Astronomy Programme, Indian Institute of Science, Bangalore 560 012, email: \texttt{prateek@physics.iisc.ernet.in}} }

\pubyear{2013}
\volume{00}
\pagerange{\pageref{firstpage}--\pageref{lastpage}}

\date{Received --- ; accepted ---}

\maketitle

\label{firstpage}

\begin{abstract}
Coronae exist in most astrophysical objects: stars, accretion disks, and individual galaxies and clusters of galaxies. 
Coronae in these varied systems have some common properties: 1) hydrostatic equilibrium in background 
gravity is a good assumption; 2) they are optically thin, i.e., photons escape as soon as they are born; 3) they have 
cooling times shorter than their ages, and thus require heating for sustenance. Generally the coronal heating mechanisms 
are quite complex but the structure of the corona is tightly constrained by the interplay of cooling, {\em global} heating, and 
background gravity. We briefly summarize the results from our studies of the intracluster medium (the cluster corona) and
draw inferences which should apply to most astrophysical coronae. 
\end{abstract}


\section{Introduction}\label{sec:intro}

Galaxy clusters are the most massive gravitationally relaxed structures in the universe. Their gravitational potential is
dominated by dark matter, and baryons only form $\approx 17\%$ of the total mass. Majority of these baryons are in form 
of the hot plasma ($10^7-10^8$ K) known as the intracluster medium (ICM). Multi-wavelength observations 
in the last decade and a half, particularly in X-rays, have revolutionized our understanding of the ICM. For brevity, we do
not include an extensive discussion of observations and various theoretical paradigms; see \cite{fab12} for a recent review. 

The focus of this paper is to review the interplay of cooling, heating, and gravity in the cores of galaxy clusters. Since these
processes are common in most astrophysical coronae, we expect the lessons learned to carry over to all astrophysical coronae such
as the solar corona and accretion disk coronae.

The cooling time in cores of some galaxy clusters (known as cool-core clusters) is much shorter than the Hubble time. In absence 
of heating, the core is expected to cool and shine brightly in soft X-rays, and the cooling gas is expected to form stars at a stupendous 
rate. Multi-wavelength observations have ruled out cooling and star-formation at the expected rate. The radio bubbles and 
X-ray cavities driven by the central supermassive black holes are powerful enough to offset catastrophic cooling in the core. The 
occurrence of cavities and bubbles in cool-core clusters has led to a picture of the ICM core in which core cooling is roughly 
balanced by mechanical heating due to the active galactic nucleus (AGN; see \cite{mcn12}).

We have recently carried out idealized numerical simulations of cluster cores in thermal balance (\cite{mcc12},\cite{sha12}). Although 
cluster cores are in rough {\em global} thermal balance, they are locally thermally unstable. We find that the ratio of the thermal instability 
timescale and the free-fall time ($t_{\rm TI}/t_{\rm ff} \approx t_{\rm cool}/t_{\rm ff}$ if heating rate is constant per unit volume) determines 
if local thermal instability can lead to extended multiphase filaments. For a large $t_{\rm cool}/t_{\rm ff}$ the overdense blobs 
respond to gravity as they are cooling; relative 
motion driven by infall leads to the mixing of blobs before they can cool to thermally stable low temperatures ($\sim 10^4$ K). If 
$t_{\rm cool}/t_{\rm ff} \lesssim 10$ the overdense blobs are able to cool to low temperatures before they can be disrupted by shear. 
This criterion for the existence of multiphase gas explains the observations of cluster cores quantitatively (Fig. 11 in \cite{mcc12}). This criterion is
quite robust and applies even with a more realistic jet heating \cite{gas12}.

The success of the thermal balance model in explaining several observations of gas in cluster and group halos motivates the application of
similar models to other astrophysical coronae. The solar corona is by far the best observed example, but even here the precise heating mechanism
is unknown. Similarly, accretion disk coronae, although unresolved, are often invoked in order to model hard spectra of AGN and X-ray binaries.
In most coronae the cooling time is quite short. Long-lived coronae with short cooling times (compared to their age) can only be sustained with coronal heating 
which  prevents the gas from condensing. Therefore, the conditions in most astrophysical coronae are expected to be very similar to those in 
the ICM, and the interplay of cooling, heating and background gravity is expected to play an important role. In the following sections we
discuss the implications of our $t_{\rm cool}/t_{\rm ff}$ criterion for various astrophysical coronae.

\section{Astrophysical Coronae}\label{sec:coronae}

\subsection{The Solar Corona}

Beyond the solar photosphere is the chromosphere with $\sim 10^4 K$ plasma. A little farther out is the corona, where the plasma temperature 
suddenly rises (across the transition region) to $\sim 10^6$ K and the density drops by a similar factor ($\sim 100$). The virial temperature at the base 
of the corona is $ T_v \equiv G M_\odot m_p/(k_B R_\odot) \sim 2 \times 10^7 $ K, about ten times larger than the coronal temperature.\footnote{The 
temperature at the coronal base can be smaller than the virial temperature because the density gradient is quite steep.} 
The coronal base in quiescent conditions is hot, subsonic, and in hydrostatic equilibrium. The solar wind is launched subsonically at the base of 
the corona. It passes through a sonic point and becomes supersonic beyond 10s of solar radii. The structure of the lower corona, and
hence the solar wind, depends on how the corona is mass loaded and how it is heated, which involve complex magnetic processes such as 
reconnection and MHD turbulence. Despite this, we believe that the lower corona is governed by the interplay of local thermal instability 
and gravity.

In steady state the lower corona is in hydrostatic and thermal balance. The outflow (which eventually becomes the solar wind) is quite subsonic and the rate 
of heating (via magnetic processes)
is balanced by radiative cooling and adiabatic expansion. These conditions (i.e., hydrostatic and thermal balance) are quite similar to conditions in cluster cores.
The cooling time and the expansion time ($r/v_r$) are longer than the free-fall time. Thus, as in cluster cores we expect multiphase gas to condense in the lower 
corona if the ratio of the cooling time to the free-fall time ($t_{\rm cool}/t_{\rm ff}$) is smaller than a critical value (we choose this to be 10 guided by cluster simulations 
but the exact value must be determined by detailed simulations). 

\begin{figure}
\centering
\includegraphics[scale=0.5]{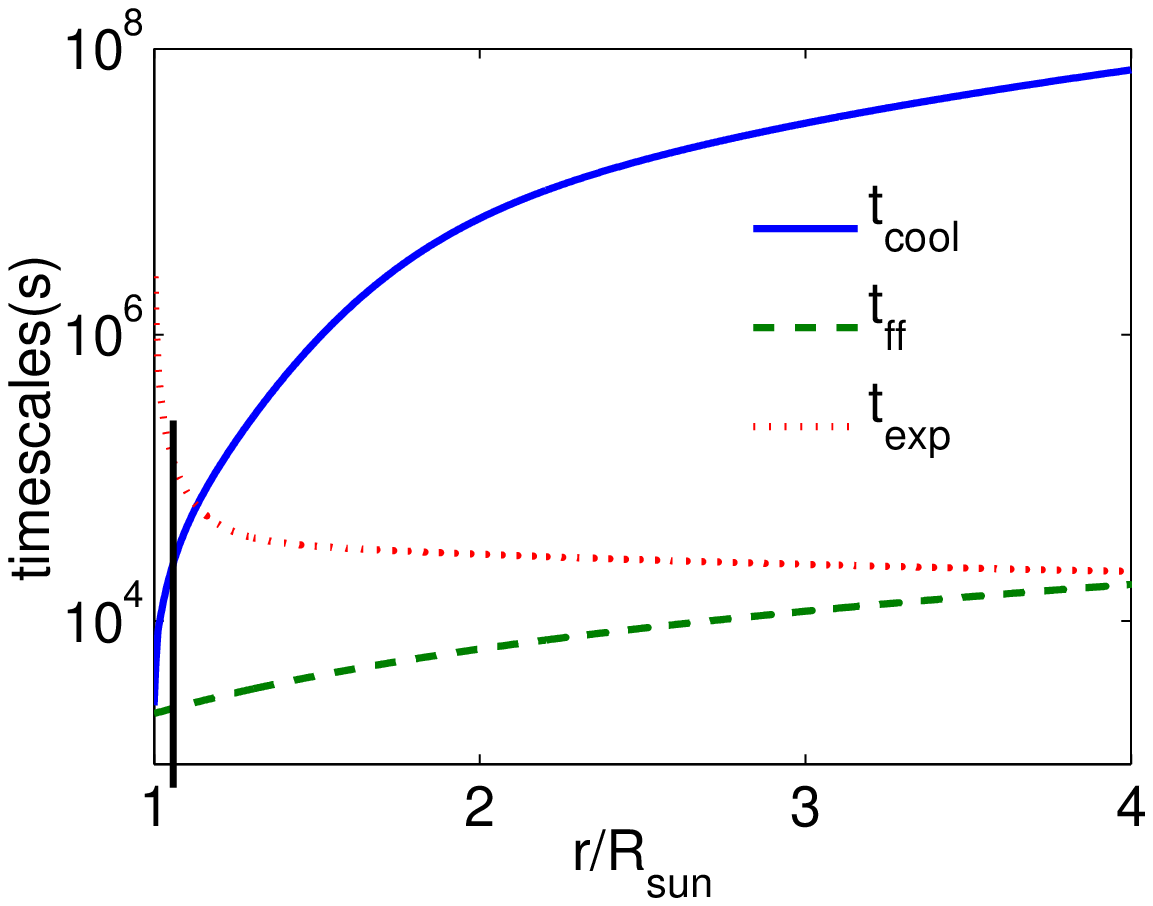}\includegraphics[scale=0.48]{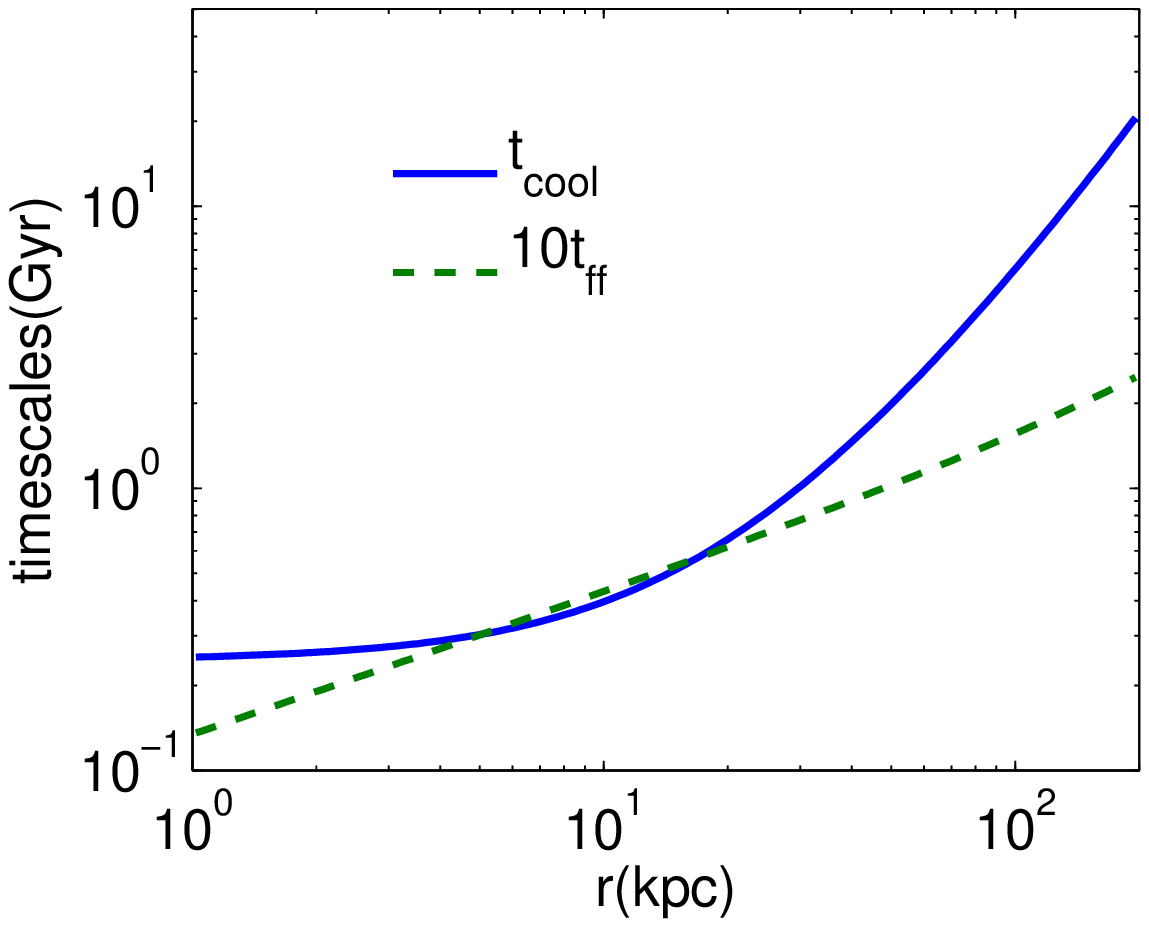}
\caption{{\em Left:}The cooling ($3/2nk_BT/n_en_i\Lambda(T)$), free-fall ($\sqrt{2r^3/GM_\odot}$) and expansion ($r/v_r$) timescales as a function of radius in the lower solar corona. The vertical line shows the radius ($\approx 1.06 R_\odot$) where $t_{\rm cool} \approx 10 t_{\rm ff}$, below which we expect multiphase gas to condense. {\em Right:} The cooling time and $10t_{\rm ff}$ for the ICM of a cool-core cluster with a core entropy of 10 keV cm$^2$. While the length and time scales in the two systems vary by several orders of magnitude, $t_{\rm cool}/t_{\rm ff}$ are in a similar range.}
\label{fig:tscales}
\end{figure}

The electron number density and temperature profiles as a function of radius (measured from the center of the sun) are roughly fit by (from $1 R_\odot$ to $4 R_\odot$;
 these fits are obtained from \cite{wit88}),
$$
n_e = 3\times 10^{8} \left (\frac{r}{R_\odot} \right )^{-10} + 5 \times 10^6 \left (\frac{r}{R_\odot} \right )^{-3} {\rm cm}^{-3},~
T = 1.7 \times 10^6 \tanh \left [10 \left ( \frac{r}{R_\odot }- 0.975 \right ) \right ] \left ( \frac{r}{R_\odot} \right) ^{-0.1} {\rm K}.
$$ 
We define the free-fall time as $t_{\rm ff } \equiv \sqrt{2r^3/GM_\odot}$. Using the cooling function for solar metallicity and the fit for the radial velocity from \cite{wit88}, in Figure \ref{fig:tscales} we plot $t_{\rm cool},~t_{\rm ff}$ and $t_{\rm exp}=r/v_r$ as a function of 
radius in the lower corona.  The timescales are of order few thousand seconds at the base of the corona and the $t_{\rm cool}/t_{\rm ff}$ ratio becomes $\lesssim 10$ in the lowest corona, which we expect to be susceptible to forming cooling multiphase gas. Coronal rain is a manifestation of the condensation of cold plasma in
 the lower corona and its eventual infall towards the chromosphere (e.g., \cite{ant10}).\footnote{For a beautiful movie see: \url{http://apod.nasa.gov/apod/ap130226.html}} Coronal rain is observed only in active regions  in the lowest corona because that is where mass loading and the density of the corona is maximum and $t_{\rm cool} \lesssim 10 t_{\rm ff}$.

All sun-like stars are expected to have thermally driven stellar winds and insights from the thermal instability model should help us interpret observations 
such as the correlation between X-ray luminosity and the mass loss rate in nearby sun-like stars (e.g., \cite{woo02}). For more X-ray active stars both heating and mass loading 
at the coronal base are higher and the coronal density is not expected to fall as steeply as in the sun; $t_{\rm cool}/t_{\rm ff} \approx 10$ is expected to happen at larger radii. The mass outflow rate $\dot{M} = 4 \pi r^2 \rho v_r$ can be approximated by using a density corresponding to the critical criterion ($t_{\rm cool}/t_{\rm ff}\approx 10$); thus, $\dot{M} \sim  r^2 m_p v_r (k_B T/\Lambda[T]) (GM/r^3)^{1/2} \propto r^{1/2} (T^{3/2}/\Lambda[T]) (v_r/c_s)$, which increases quite rapidly with temperature. Therefore, the coronae of X-ray active stars are expected to be hotter, with shallower density profile, and hence with a higher mass outflow rate. However, there is an upper limit on the mass loading of thermally driven winds because temperature cannot be much larger than the virial temperature and the Mach number in the corona ($v_r/c_s$) must be $< 1$. This can explain the observed breakdown of the roughly linear scaling of the stellar mass outflow rate and the X-ray luminosity. We plan to investigate these assertions  more rigorously in future.

\subsection{Accretion Disk Coronae}
Coronae are important components of accretion flows. Most accretion flows show hard radiation which requires a hot corona which can radiate either via free-free emission or via inverse Compton scattering of soft thin disk photons (e.g., \cite{cha95}). A thin disk condenses from the hot accretion flow only when the infall (viscous) time is longer than the cooling time; i.e., if the flow can cool before it is accreted. However, once a thin disk forms, the structure of the corona sandwiching the cold thin disk is governed by the interplay of local thermal instability and gravity. The corona, which is heated by viscous stresses and  magnetic processes  and cooled by optically thin radiation is in rough thermal and hydrostatic balance. If an overdense blob condenses because of a small $t_{\rm cool}/t_{\rm ff}$, it will fall vertically onto the cold disk on a free-fall timescale; the blob does not fall radially because it is supported by rotation (and not thermal pressure) in that direction. Our numerical simulations \cite{das13} show that the lower corona has a density upper limit corresponding to $t_{\rm cool}/t_{\rm ff} \gtrsim 10-100$, irrespective of viscosity and the mass accretion rate. For large accretion rates more and more mass condenses into an optically thick, geometrically thin cold disk but the maximum density of the corona is still governed by $t_{\rm cool}/t_{\rm ff} \gtrsim 10-100$. For details see our recently submitted paper (\cite{das13}). 

Both the stellar and accretion disk coronae are different from the ICM in that $t_{\rm cool}/t_{\rm ff}$ is smallest at the base (Fig. \ref{fig:tscales}). Therefore, cold gas condenses close to the interface between the lower corona and the chromosphere/photosphere. Cool-core clusters, however, generally have a minimum in $t_{\rm cool}/t_{\rm ff}$ at 10s of kpc  (Fig. \ref{fig:tscales}) which results in the condensation of {\em extended} mutiphase gas filaments. While the details of astrophysical coronae depend on the system, we expect the physics of local thermal instability and its interaction with background gravity to be robust and applicable to most astrophysical coronae.

\label{lastpage}


\end{document}